\documentclass{jetp} 

\begin{document}
\English

\title{Phenomenological Extension for Tidal Charge Black Hole}

\setaffiliation1{Sternberg Astronomical Institute, Lomonosov Moscow State University, Universitetskii Prospekt, 13, Moscow, 119234, Russia}

\setaffiliation2{Department of Quantum Theory and High Energy Physics, Physics Faculty, Lomonosov Moscow State University, Vorobievi Gory, 1/2, Moscow, 119234, Russia}

\setaffiliation3{Department of Physics and Astronomy, University of Sussex, Brighton, BN1 9QH, United Kingdom}
\setaffiliation4{Dubna State University, Universitetskaya St. 19, Dubna 141982, Russia}

\setaffiliation5{Department of Astrophysics and Stellar Astronomy, Physics Faculty, Lomonosov Moscow State University, Vorobievi Gory, 1/2, Moscow, 119234, Russia }

\setauthor{S.~O.}{Alexeyev}{12}
\setauthor{B.~N.}{Latosh}{34}
\setauthor{V.~A.}{Prokopov}{15}
\setauthor{E.~D.}{Emtsova}{15}

\abstract{
A simple phenomenological extension of the black hole solution with tidal charge is proposed. Empirical data on the Sgr A* is consistent with the suggested metric which serves as a generalisation of the Reissner-Nordstr\"om one. Such a generalisation includes the leading effects beyond general relativity so, the discussed metric can explain wider range of gravitational effects. We discuss physical features of an object described by the proposed metric, namely, the size of its shadow and the innermost stable circular orbit radius.
}

\maketitle

\section{Introduction}

General Relativity (GR) is admitted to be the best theory of gravity, as it provides a correct description of multiple gravitational phenomena \cite{Will:2014kxa,TheLIGOScientific:2016src}. At the same time the existence of dark matter and dark energy provides a ground to consider GR as a relevant theory only on small spacial scales \cite{Weinberg:1988cp,Clowe:2006eq,Ade:2015xua}. Basic physical principles that may be used to modify GR are well-understood and widely implemented in so-called extended gravity models \cite{Capozziello:2011et,Berti:2015itd}. The spectrum of these models includes multiple $f(R)$ gravity and scalar-tensor models, including Horndeski models \cite{Sotiriou:2008rp,DeFelice:2010aj,Charmousis:2011bf}.

A particular modify gravity model can be considered as a suitable generalization of GR, if it provides a better description of gravitational phenomena at least at one spacial scale. Because of this feature one is only interested in modified gravity models that passes Solar system tests, for instance, such a model should have post-Newtonian parameters (and post-Keplerian parameters for binary pulsars) consistent with the current empirical data \cite{Will:2014kxa,Berti:2015itd,Dyadina:2018ryl}.

Within GR a non-rotating black hole without electric charge is described by the Schwarzschild metric. At the same time recent results \cite{Zakharov:2014lqa} show that the Reissner-Nordstr\"om metric with non-vanishing charge is consistent with the empirical data from Sgr A* black hole \cite{BinNun:2009jr,BinNun:2010se,BinNun:2010ty}. There are strong theoretical evidences that a real ``astrophysical'' black hole cannot have a significant electric charge. At the same time modified gravity models, namely, Randall-Sundrum one, predict that a metric similar to the Reissner-Nordstrom one serves as an exact solution of corresponding field equations \cite{Dadhich:2000am}. Such a solution describes a black hole without an electric charge, but with an additional parameter called ``tidal charge'' which is caused by the gravitational field propagating in additional dimensions (bulk). At the same time the physical mass of a black hole formed on the brane does not depend on the additional dimensions, so the corresponding solution has the same physical mass as in the Reissner-Nordstr\"om case \cite{Alexeyev:2015mta}. Moreover, the contribution of the tidal charge may have an opposite sign in contrast to the Reissner-Nordstrom case. This fact helps one to distinct it from the electric charge. For the sake of simplicity we refer to the metric \cite{Dadhich:2000am} as to the Reissner-Nordstrom metric with a tidal charge following the notations from \cite{Zakharov:2014lqa}.

As the empirical data is consistent with the Reissner-Nordstrom metric with a tidal charge then there is a ground to assume that the external gravitational field of a real black hole may be described by some generalization of this metric. So we propose a generalization of the Reissner-Nordstrom metric that accounts for more fine modified gravity effects which fall faster that $r^{-2}$. The corresponding effects will be detected in the nearest future via projects aimed on measuring black hole shadow properties \cite{Fish:2010wu,Zakharov:2018awx}. So the main scope of this paper is to provide an additional way to connect observational data with new metric parameters. Namely, we study possible influence of new modified gravity effect on the innermost stable circular orbit (ISCO) radius and the black hole shadow.

The paper is organised as follows. In part 2 we introduce the modified metric and discuss its properties; part 3 is devoted to properties of the shadows cast by such an object; in part 4 ISCO is discussed, and part 5 contains our conclusions.

\section{Phenomenological metric}

The simplest ansatz for a static spherically-symmetric metric is (in the Planck units $G=c=\hbar=1$):
\begin{equation}\label{sim}
ds^2=\Delta(r) dt^2 - \cfrac{dr^2}{\Delta(r)} - r^2 (d\theta^2+ \sin^2\theta d\phi^2 ).
\end{equation}
A particular form of function $\Delta$ is defined by a particular physical setting. In the case of the Reissner-Nordstr\"om black hole function $\Delta$ reads:
\begin{align}
  \Delta = 1 - \cfrac{2M}{r}+\cfrac{Q^2_\text{electric}}{r^2}.
\end{align}
In case of the black hole with a tidal charge \cite{Dadhich:2000am} function $\Delta$ has a similar form:
\begin{align}
  \Delta = 1 -\cfrac{2M}{r} - \cfrac{Q_\text{tidal}}{r^2}.
\end{align}
The $Q_\text{tidal}$ sign can be both negative or positive. In such a way all possible modified gravity effects that falls as $r^{-2}$ cover by these two cases. Therefore the only possible way to account for new gravitational effects is to introduce a new term in $\Delta$ that falls as $r^{-3}$. In this paper we use the following metric function $\Delta$:
\begin{equation}\label{nm1}
  \Delta(\hat{r})=1-\cfrac{2}{\hat{r}}+\cfrac{q}{\hat{r}^2}+ \cfrac{\alpha}{\hat{r}^3} ~.
\end{equation}
In this notation we use length units normalised by the black hole mass, namely $\hat{r}=r/M$; quantities $q$ and $\alpha$ are free dimensionless parameters. It should be noted, that such a metric can be considered as Taylor series in $\hat{r}^{-1}=M/r$, so one relates any given modified gravity metric with \eqref{nm1} via Taylor series. It is crucial to highlight, that such a generalization is not an exact solution of any particular field equations and we use it as a phenomenological ansatz. For the time being we treat this metric as an exact solution and search for a way to relate $M$, $q$, and $\alpha$ with observed and measured quantities, for example, the size of a black hole shadow. In such a way the metric \eqref{nm1} serves as a toy model which parameters can be related both with empirical data and fundamental modified gravity model parameters.

\begin{figure}[!hbtp]
\begin{center}
\includegraphics[width=8cm,height=5cm]{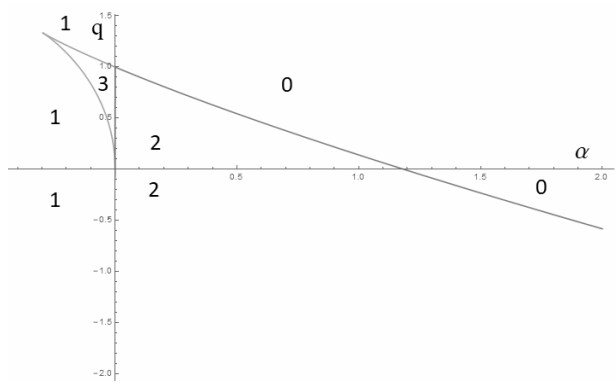}
\end{center}
\caption{Configuration space of the model with the third correction divided in areas with various number of horizons.}
\label{hor2}
\end{figure}

In order to implement such a logic one should define the area of applicability of the presented metric. As we consider it as an expansion in $\hat{r}^{-1}$ it should not be used in small $r$ area. The metric has three free parameters and they all have an influence on the corresponding area of applicability. In order to establish a precise constraints one should find the event horizon radius, as it marks the area where gravity enters the strong field regime. The horizon position is defined by the equation $\Delta(\hat{r}_h)=0$. As function \eqref{nm1} is a third order polynomial, the equation $\Delta(\hat{r})=0$ have up to three real roots, so such an object can develop up to three horizons. In Fig. \ref{hor2} we present a phase diagram that shows the number of horizons that the metric can have. The upper curve corresponds to the following equation
\begin{equation}\label{Horizon_shift_curve}
    \alpha=\cfrac{2}{27} (\sqrt{4-3q}-1 )(\sqrt{4-3q}+2)^2.
\end{equation}
The equation for the lower curve in Fig \ref{hor2} reads
\begin{equation}
    \alpha=-\cfrac{2}{27}(\sqrt{4-3q}+1 )(\sqrt{4-3q}-2)^2.
\end{equation}
 
\begin{figure*}[t] 
\begin{center}
\includegraphics[width=\textwidth, height=5cm]{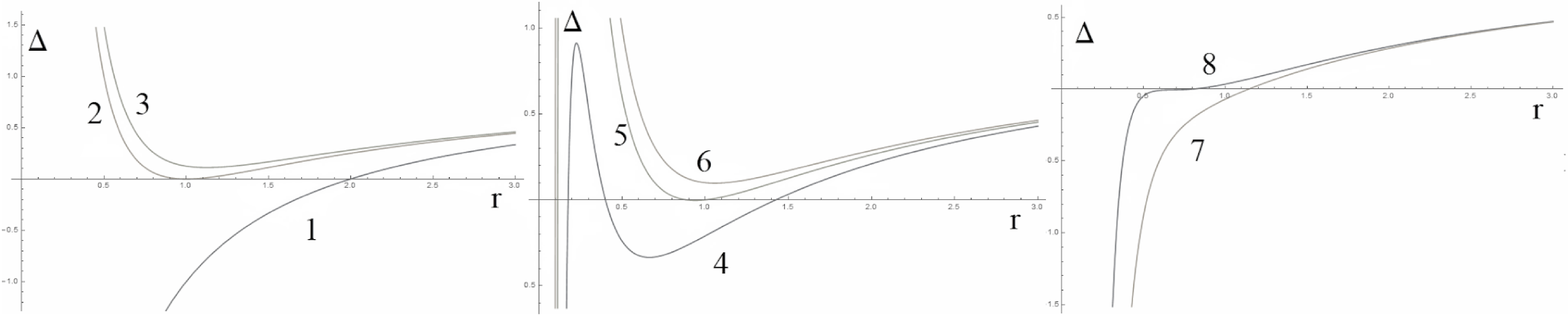}
\end{center}
\caption{Plots of the metric function $\Delta$ of the generalised Reissner-Nordstr\"om ansatz with different values of $q$ and $\alpha$. 1)$\alpha=0$, $q=0$; Schwarzschild BH 2) $\alpha=0$, $q=1$; Reissner-Nordstrom BH with critical charge. 3) $\alpha=0$, $q=9/8$; "naked singularity" in Reissner-Nordstrom ansatz. 4) $\alpha=-0.1$, $q=0.89$; BH with three horizons. 5) $\alpha=-0.1$, $q=1.1$; BH with two horizons. 6) $\alpha=-0.1$, $q=1.2$; BH with one compact horizon. 7) $\alpha=-0.4$, $q=1.33$; BH with one horizon. 8)  $\alpha=-0.296$, $q=1.33$: end of horizon disappearance curve. }
\label{31}
\end{figure*}

Unlike the standard Reissner-Nordstr\"om case, the function \eqref{nm1} has a new external horizon. If such a metric would describe an exact solution in a modified gravity model, than an external observer would detect the finite shift of the BH radius. For the sake of simplicity we call this phenomenon as {\it horizon shift}. Curves number $2$-$6$ in Fig. \ref{31} illustrate the way by which metric \eqref{nm1} forms the new external horizon. Configurations corresponding to the horizon shift lie on the curve \eqref{Horizon_shift_curve}. Plots demonstrating these critical points are presented in Fig. \ref{hor2} and Fig.\ref{res1}. At the same time (see Fig. \ref{hor2}) the condition of a horizon disappearing includes the vanishing of the metric function $\Delta$ and its first derivative:
\begin{equation}\label{horizon4}
  \Delta(\hat{r}) = 0, \qquad \cfrac{d\Delta(\hat{r})}{d\hat{r}} = 0 .
\end{equation}  

An additional comment on the applicability of the model is due. We treat the generalised Reissner-Nordstrom metric as first terms of perturbation series to fit the empirical data. In other words, the metric should serve as a proper fit for a gravitational field around a real BH. So this metric may not be valid for the BH internal structure. The external horizon of the metric \eqref{nm1} marks the area where the generalised Reissner-Nordstrom metric enters strong gravity regime. As the used approximation cannot be a proper ansatz in the strong gravity regime, the external horizon naturally defines the area of applicability of the metric. If there is no horizon at all, then the only natural cut-off conditions read $M\ll r$, $q \ll 1$, $\alpha \ll 1$. Such a premise allows one to consider the area covered by the event horizon irrelevant and study particle motion only outside the horizons.

\section{Shadow properties}

A BH forms a shadow by capturing photons travelling close to the horizon. A photon radiated by a distant source either escapes a black hole and reaches a distant observer, or been captured by a black hole. A distant observer can only detect photons that escaped a black hole, so an Earth telescope can only detect a bright spot formed by photons that were not captured by a black hole. Photons captured by a black hole form a dark spot which is called a black hole shadow. The size and the form of the shadow is defined by black hole parameters. It should also be noted, that the position of an observer and the orientation of the black hole spin also affect the shape of a black hole shadow. The influence of the black hole spin is irrelevant for our study, as we consider a black hole with vanishing spin. And we implement a method of a black hole shadow size study which allows one to obtain results independent on the distance to an observer. In such a way the main goal of this section is to study the black hole shadow of the proposed metric.

The critical state of photons, when they are neither captured by a black hole, nor free to escape to spacial infinity is called a photon sphere. The radius of the photon sphere defines a characteristic spacial scale of black hole shadow formation. Within GR the Schwarzschild solution should be used to establish such a relation. It should be noted, that one can also use other solutions to evaluate the same scale, namely, the Schwarzschild-de Sitter solution. We prefer to not to use it, as this metric account for the global structure of the Universe, which lies beyond the scope of this paper. For the Schwarzschild metric the radius of the photon sphere is $3/2$ of the black hole radius. The metric we propose use $M/r$ as a small parameter which modules small corrections to the Reissner-Nordstrom metric. For the GR case the value of $M/r$ of the photon sphere is $2/3$ which place the formation of the black hole shadow on the border of the model area of applicability. We want to show that the existence of new parameters $q$ and $\alpha$ can decrease the value of the photon sphere radius. Therefore one should expect, that the existence of new gravitational effects can change the value of the photon sphere radius thereby improving the convergence properties of the series and extending the area of applicability of the correspondent metric.

In order to calculate the shadow size it is necessary to resolve a scattering problem. Namely, one should find a photon gravitational capture cross-section. It is convenient to use the photon impact parameter $D$ (in the black hole mass unites) as the key variable. Photons with small $D$ are captured by the object and cannot reach a distant observer, while photons with large $D$ avoid a black hole and travel further. The size of the shadow is defined by the value of the critical impact parameter $D_c$ corresponding to photons occupying the photon sphere. Usually the angular size of the dark spot generated by a black hole is called the black hole shadow. In this case, as we highlighted before, the size of the shadow depends the distance between an observer and the black hole. Therefore the size of the shadow can only be calculated, if the distance to the black holes is known. In order to avoid this limitation and to obtain a shadow description that does not depend on the distance to an observer, we study the linear size of the shadow, namely, the value of the critical impact parameter $D_c$. In such an approach the angular size of a black hole shadow can be recovered, if the linear size $D_c$ is divided by the linear distance to a black hole. It also should be highlighted, that such a method only applicable for asymptotically flat metrics. Such an approach is suitable for our purposes, as the contemporary precision level of observational apparatus can only resolve shadows of the nearest black holes. In such a way, despite the fact, that in the most general case an accelerated expansion of the Universe should be taken into account, on the practical ground its influence on objects that can be studied is negligibly small.

The case $\alpha=0$ corresponds to Reissner-Nordstr\"om metric and it was studied in \cite{Zakharov:2014lqa}. It was shown that Reissner-Nordstr\"om metric with $q>9/8$ has no shadow at all. To be exact, the shadow size does not vary smoothly with $q$. The critical value of the impact parameter $D$ decreases with $q$ down to $q = 9/8$ where $D_c\simeq 3.674$. The the size of the shadow experience a finite shift down to zero and the shadow disappears.

It also should be noted, that Reissner-Nordstrom metric with $q>1$ describe a {\it naked singularity}. Such a metric appears to be an exact solution within a particular Randall-Sundrum model \cite{Dadhich:2000am}, so it still can be considered as a real object without a horizon, but with non-vanishing shadow size. Withing our approach there is no room for such an interpretation, because parameter $q$ is not small. Therefore we interpret this result as a fact, that metric \eqref{nm1} can only describe object with shadow size bigger than $D=4$.

Photon geodesic equation reads:
\begin{align}
  & \left(\cfrac{d \hat{r}}{d\tau}  \right)^2 + \left(1 - \cfrac{2}{\hat{r}} + \cfrac{q}{\hat{r}^2} + \cfrac{\alpha}{\hat{r}^3}\right) \cfrac{L^2}{\hat{r}^2} = E^2, \label{rt}   \\
  & \cfrac{d\phi}{d\tau} = \cfrac{L}{\hat{r}^2}, \label{ft}
\end{align}
here $E$ is the photon energy, $L$ is its angular momentum, and $\tau$ is an affine parameter. We map photons motion by the angle $\phi$, so the equation reads:
\begin{equation}\label{rf}
u(r) = \left(\cfrac{d \hat{r}}{d \phi}\right)^2 = \cfrac{\hat{r}^4}{D^2} - \hat{r}^2\left(1 - \cfrac{2}{\hat{r}} + \cfrac{q}{\hat{r}^2} + \cfrac{\alpha}{\hat{r}^3}\right),
\end{equation}
where $D = L/E$ is the photon impact parameter. The value of $u = (d\hat{r}/d\phi)^2$ is always positive, so motion is possible if and only if the right-hand side of \eqref{rf} is positive. 

\begin{figure}[!hbtp]
\begin{center}
\includegraphics[width=8cm,height=5cm]{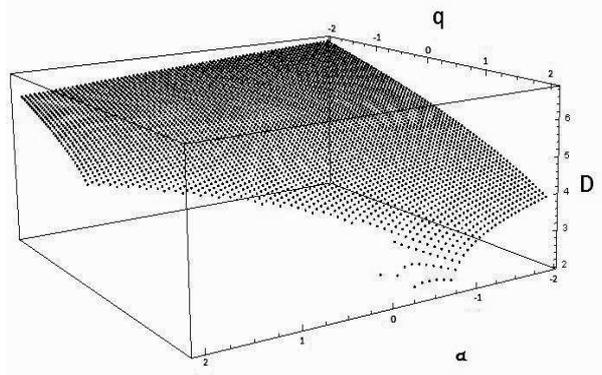}
\end{center}
\caption{The dependence of the critical impact parameter $D$ on $q$ and $\alpha$}
\label{3d}
\end{figure}

The radius of the photon sphere is given by the following:
\begin{eqnarray}\label{ss}
 u(r)=0 , & \cfrac{du(r)}{dr}=0.
\end{eqnarray}
The system \eqref{ss} was solved numerically and the corresponding plot for the critical impact parameter is presented in Fig. \ref{3d}. 

System \eqref{ss} has no solutions for certain configurations of metric parameters. In full analogy with the standard Reissner-Nordstrom metric, the absence of solution for \eqref{ss} indicates the absence of the shadow.

\begin{figure}[!hbtp]
\begin{center}
\includegraphics[width=8cm]{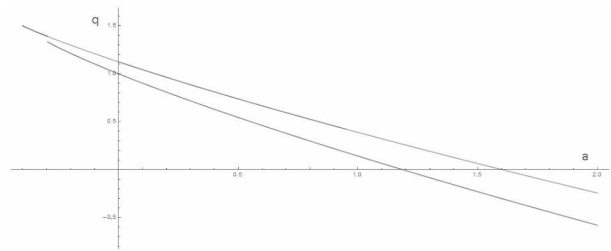}
\end{center}
\caption{Critical curves corresponding to finite shift of the horizon radius and to the finite shift of size of the shadow. The upper curve corresponds to the shadow size shift, while the lower one to the horizon radius shift.}
\label{res1}
\end{figure}

Configurations with and without a shadow are separated by a curve defined by the solution of the following equations:
\begin{eqnarray}\label{css}
u(r)=0 , & \cfrac{du(r)}{dr}=0, &\cfrac{d^2 u(r)}{dr^2}=0.
\end{eqnarray}
Its solution was obtained numerically and presented in Fig. \ref{res1}. 
Following the approach from \cite{Chandrasekhar:1985kt} we find the critical value of the impact parameter $D_{orb} = D_{orb}(q,\alpha)$. In the Schwarzschild case ($\alpha = q = 0 $) the critical impact parameter value is calculated explicitly: $D_{orb} = 3 \sqrt{3}$ \cite{Chandrasekhar:1985kt}. For $\alpha = 0$ the case is reduced to the standard Reissner-Nordstrom ansatz \cite{Zakharov:2014lqa}. 

The dependence of the critical impact parameter $D$ on $q$ and $\alpha$ is presented in Fig. \ref{3d}. As it was noted earlier, the metric can experience a finite shift of the event horizon radius. A similar effect realizes for the shadow, as its size can also experience a finite shift with an infinitely small variation of metric parameters. The plot showing corresponding critical curves in the configuration space is presented in Fig \ref{res1}. Curves presented in Fig. \ref{res1} do not intersect each other and can be extended infinitely only in one direction.

For positive values of $\alpha$ the shadow disappears when the object has no horizon. For ($-0.296 < \alpha <0.00$) the shadow experiences a finite shift, while the object has one horizon. BHs with ($\alpha < -0.296$) have only one horizon. 

It is important to emphasis that a one-to-one correspondence between the shadow size and metric functions does not takes place due to the fact that the metric has two independent parameters $q$ and $\alpha$. Therefore it is not possible to use one data source, such as the size of a BH shadow, to establish both $q$ and $\alpha$.

\begin{figure}[!hbtp]
\begin{center}
\includegraphics[width=8cm]{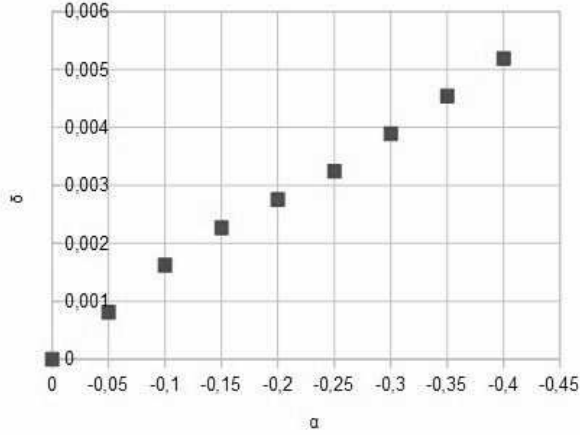}
\end{center}
\caption{Deviation parameter $\delta=|D(\alpha)-D(0)|/D(0)$ for the BH with same shadow size corresponding to $D_{sh}=3\sqrt{3}$.}
\label{delta}
\end{figure}

Gravitational lensing can provide an additional information on metric functions, as BHs with the same shadow size have different external field and act as different gravitational lenses. In such a way one has to solve equation \eqref{rf} to extract the information about the strong lensing. We obtained the dependence between the deflection angle, BH parameters ($q,\alpha$), and the position of a star on the image plane (which is parametrized by the impact parameter) numerically. As a result two BHs with equal shadow sizes have different value of the parameter $\alpha$. Therefore there is an opportunity to describe configurations with any given shadow size by $\alpha$. We take the case $\alpha = 0$ as a reference point to see how the position $D$ (the impact parameter) of a distant star at the image plain changes depending on $\alpha$. In Fig. \ref{delta} we present a plot of the deviation parameter $\delta$ for an object which shadow size corresponds to impact parameter $D_{sh}=3\sqrt{3}$ and deflection angle of the light source $\phi=\pi/2$. The deviation parameter is defined by the following:
\begin{equation}
\delta=\cfrac{\lvert D(\alpha)-D(0)\rvert}{D(0)}.
\end{equation} 
The Fig. \ref{delta} shows that the correction to the critical impact parameter induced by $\alpha$ is by three orders of magnitude smaller that the value itself. The same ratio is valid for other values of $D_{sh}$. Therefore the evaluation of $\alpha$ by gravitational lensing requires precise measurements. 

\section{The innermost stable circular orbit radius}

The ISCO radius defines a characteristic length scale of orbital motion. As no circular motion is possible below ISCO, it serves as a natural constraint on the internal radius of an accretion disk and may be constraint by direct observational data. The case $\alpha=0$ corresponds to the Reissner-Nordstr\"om metric and was studied in \cite{Pugliese:2010ps}. Within GR the ISCO radius is three times larger that the event horizon radius. This fact support the applicability of the metric we study. As we highlighted, the generalization of the Reissner-Nordstrom metric we study can be treated as a Taylor expansion in small parameter $M/r$. The value of this parameter on ISCO within GR is $1/3 \ll 1$. Therefore even within GR such an expansion has suitable convergence properties. Corrections associated with new beyond GR gravity effects can influence the value of the ISCO radius and change correspondent convergent properties. However, it is required to find the ISCO radius first in order to evaluate its influence on the model features.

Study of the ISCO radius follows the well-known approach \cite{Chandrasekhar:1985kt}. Equation of motion of a test particle with a unit mass reads:
\begin{eqnarray}
&& \left(\cfrac{d \hat{r}}{d \tau} \right)^2 + U = E^2, \label{rt1} \\
&& \cfrac{d\phi}{d\tau} = \cfrac{L}{\hat{r}^2} \label{ft1} \\
&& U = \left(1-\cfrac{2}{\hat{r}}+\cfrac{q}{\hat{r}^2}+\cfrac{\alpha}{\hat{r}^3}\right)\left(1+\cfrac{L^2}{\hat{r}^2}\right). \label{rt2}
\end{eqnarray}
where $E$ is the energy of a particle, $L$ is particle angular momentum, and $U$ is the potential energy. Motion is possible if $E^2 \geq U$. The orbit is circular if its radius is static: $\cfrac{d\hat{r}}{d\tau} = 0$ which is equivalent to the equation $U(\hat{r}_{circ}, L_{circ}) = E^2$. The orbit is stable if $U(\hat{r}_{circ}, L_{circ})$ is a local minimum. ISCO is situated at the inflection point of the potential energy, so it can be found from the equation:
\begin{eqnarray}\label{eq555}
\cfrac{ dU}{dr}=0 , \qquad \cfrac{d^2U}{dr^2}=0 .
\end{eqnarray}

For certain values of parameters $q$ and $\alpha$ the metric admits more than one solution of \eqref{eq555}. In that case a stable orbit with the biggest radius considered to be the last stable one. This is true because, as we highlighted, we treat the generalised Reissner-Nordstrom metric only as the ansatz describing external field of a real BH. So stable circular orbits with smaller radii lie outside the area of our consideration.

The metric admits exotic solutions with zero orbital momentum:
\begin{equation}
 \cfrac{ dU}{dr} = 0, \qquad L = 0.
\end{equation}
Such solutions were first described in \cite{Pugliese:2010ps}. They correspond to a particle located on some distance from the BH. In these points the total gravitational force is exactly equal to zero. Such an equilibrium is possible because of the different signs in function \eqref{nm1}. In \cite{Pugliese:2010ps} these trajectories were called {\it orbits with zero orbital momentum}. Despite the existence of such orbits, we do not consider them, as they do not affect physics of orbital motion.

\begin{figure}[!hbtp]                      
\begin{center}
\includegraphics[width=8cm]{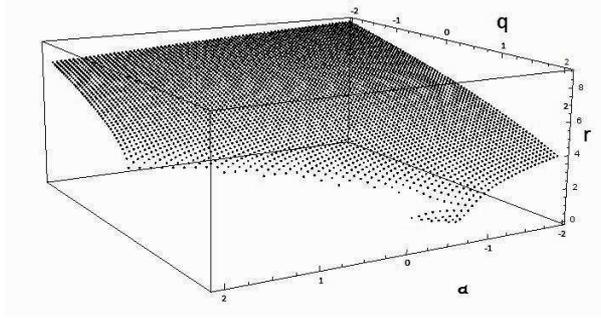}
\end{center}
\caption{The size of ISCO $r(q,\alpha)$}
\label{4d}
\end{figure}

\begin{figure}[!hbtp]                      
\begin{center}
\includegraphics[width=8cm]{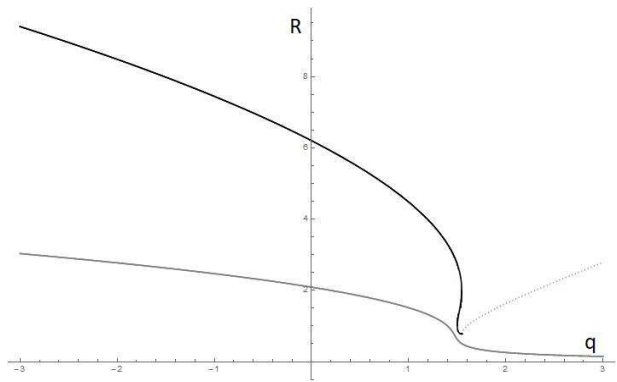}
\end{center}
\caption{The size of ISCO (upper curve) for  $\alpha = -0.4$. The lower curve corresponds to the size of the horizon. Dashed one shows to the orbits with zero moment.}
\label{orb-1}
\end{figure}

The dependence of ISCO radius on $q$ and $\alpha$ is presented in Fig. \ref{4d}. The ISCO radius dependence on $q$ for a fixed value of $\alpha=-0.4$ is presented in Fig. \ref{orb-1}.

\section{Conclusions}

In this paper we studied a generalization of the Reissner-Nordstrom metric performed by an introduction of a term proportional to $r^{-3}$. Such a generalization allows one to account for beyond GR effects that cannot be described by the pure Reissner-Nordstrom metric \cite{Zakharov:2014lqa,Pugliese:2010ps,Zakharov:2018awx}. We treat this metric as a phenomenological ansatz which allows us to relate metric parameters with observational data. At the same time one can consider such a metric as a Taylor expansion of a static spherically symmetric black hole space time in a modified gravity model. Such an approach allows one to relate parameters of a given modified gravity model with metric functions and to compare them with empirical data.

Reissner-Nordstrom ansatz (admitting both sign of $r^{-2}$ term) was considered in \cite{Zakharov:2014lqa,Pugliese:2010ps,Zakharov:2018awx}. In that paper it was shown that the BHs shadow reaches its minimal size when the charge parameter is equal to $q = 1$, which corresponds to the photon impact factor value $D=4$. At the same time the minimal ISCO radius is equal to $r_\text{ISCO} = 4$. Phase diagram of the metric is presented in Fig \ref{hor2}. Curves corresponding to horizon radius and shadow size shifts are presented in Fig \ref{res1}. 

In full analogy with the standard case the extended Reissner-Nordstrom metric describes both black holes and naked singularities with various shadow sizes. Unlike the original case, the extended metric can experience a final shift of the horizon. The similar phenomenon occurs for the shadow size, as it can experience a finite shift with infinitely small variation of metric functions. Moreover BH configurations described by the metric can have arbitrary small shadow ($D<4$) and ISCO radius ($\hat{r}<4$). The similar effect does not occur in Reissner-Nordstrom metric, as it has a smaller configuration space. If this generalization of the Reissner-Nordstrom metric is treated as Taylor series the aforementioned results allow one to establish the area of applicability of that metric. As the metric has three free parameters $M$, $q$, and $\alpha$, its area of applicability is defined by all of them, despite the fact, that the expansion is performed in $M/r$. Therefore the event horizon radius should be treated as a natural spacial scale defining the area where gravity enters the strong field regime and the metric loses its convergence features.

Here it is important to emphasis that the decrease of the shadow size could also be caused by effects taking place in plasma around a BH \cite{Perlick:2015vta,Perlick:2017fio}. Although these effects are competing, plasma effects depend on the wavelength, so they can be separated by means of observation in different spectra ranges. 

\section{Acknowledgements}

The work was supported by RFBR grant No.16-02-00682

\bibliographystyle{elsarticle-num}
\bibliography{mybibfile}

\end{document}